\documentclass[pre,aps,twocolumn,amsmath,amssymb,showpacs,showkeys,unsortedaddress]{revtex4-1}
\usepackage{graphicx}% Include figure files
\usepackage{epsfig}
\usepackage{dcolumn}% Align table columns on decimal point
\usepackage{bm}% bold math

\newcommand{\la}{\left\langle}         %% <...> (left)
\newcommand{\ra}{\right\rangle}        %% <...> (right)
\newcommand{\fg}[1]{Fig.~\ref{#1}}     %% Fig.  ...
\begin{document}
\title{Kinetic exchange models for social opinion formation}

\author{Mehdi Lallouache}
  \email{mehdi.lallouache@ens-cachan.fr}
  \affiliation{D\'epartement de Physique, \'Ecole Normale Sup\'erieure de Cachan, 94230 Cachan, France}
  \affiliation{Chaire de Finance Quantitative, Laboratoire de Math\'ematiques Appliqu\'ees aux Syst\`emes, \'Ecole Centrale Paris, 92290 Ch\^atenay-Malabry, France}

\author{Anirban Chakraborti}
  \email{anirban.chakraborti@ecp.fr}
   \affiliation{Chaire de Finance Quantitative, Laboratoire de Math\'ematiques Appliqu\'ees aux Syst\`emes, \'Ecole Centrale Paris, 92290 Ch\^atenay-Malabry, France}
   
\author{Bikas K. Chakrabarti}
 \email{bikask.chakrabarti@saha.ac.in}
 \affiliation{Centre for Applied Mathematics and Computational Science,
Saha Institute of Nuclear Physics,
1/AF Bidhannagar, Kolkata 700 064, India}
\affiliation{Economic Research Unit, Indian Statistical Institute, 203 B. T. Road, Kolkata 700 018, India}
  \affiliation{Chaire de Finance Quantitative, Laboratoire de Math\'ematiques Appliqu\'ees aux Syst\`emes, \'Ecole Centrale Paris, 92290 Ch\^atenay-Malabry, France}

\date{\today}

\begin{abstract}
We propose a minimal model for the collective dynamics of opinion formation
in the society, by modifying kinetic exchange dynamics studied in the context of
income, money or wealth distributions in a society. This model has an intriguing spontaneous symmetry
breaking transition.
\end{abstract}

\pacs{87.23.Ge 02.50.-r}
\keywords{Econophysics; Sociophysics; kinetic theory}

\maketitle

%%%%%%%%%%%%%%%%%%%%%%%%%%%%%%%%%%%%%%%%%%%%%%%%%%%%%%%%%%%%%%%%%%%%%%%%%
\section{Introduction}
%%%%%%%%%%%%%%%%%%%%%%%%%%%%%%%%%%%%%%%%%%%%%%%%%%%%%%%%%%%%%%%%%%%%%%%%%
A very interesting problem in studying society and social
dynamics is the one of ``opinion formation'', which is a collective dynamical phenomenon,
and as such are closely related to problems of competing cultures or languages \cite{Castellano2009,Axelrod1997,Abrams2003}. 
It deals with a ``measurable'' response of the society to e.g., political issues, acceptances of
innovations, etc. A number of models of competing options have been introduced
to study it, e.g., the ``voter'' model (which has a binary opinion variable with the
opinion alignment proceeding by a random choice of neighbors) \cite{Holley1975}, or the Sznajd-Weron discrete opinion formation model (where
more than just a pair of spins is associated with the decision
making procedure) \cite{Sznajd-Weron2000}. 
There have been studies of systems with more than just two possible opinions \cite{Vazquez2003}, or where the opinion of individuals is represented
by a ``continuous'' variable \cite{Hegselman2002,Deffuant2000,Fortunato2005}.
Since opinion formation in a human society is mediated by social
interactions between individuals,  such social dynamics was considered to take
place on a network of relationships by Holme and Newman \cite{Holme2006}. Several other significant studies have followed, which we not not mention here.

A two body exchange dynamics has already been developed in the context of modelling
income, money or wealth distributions in a society \cite{Yakovenko2009,Arnab2007,Patriarca2010,Chakraborti2010a}. Detailed analytical structure of
the collective dynamics in these models are now considerably well-developed \cite{Repetowicz2005,Matthes2008}.
Here, we propose a minimally modified version of those models for the collective dynamics of opinion formation
in the society.

%%%%%%%%%%%%%%%%%%%%%%%%%%%%%%%%%%%%%%%%%%%%%%%%%%%%%%%%%%%%%%%%%%%%%%%%%
\section{Kinetic exchange models of market}
%%%%%%%%%%%%%%%%%%%%%%%%%%%%%%%%%%%%%%%%%%%%%%%%%%%%%%%%%%%%%%%%%%%%%%%%%
Recently physicists and mathematicians have been interested in studying the wealth distributions in
a closed economy using kinetic exchange mechanism, which has led to new insights into this field (see Refs. \cite{Yakovenko2009,Arnab2007,Patriarca2010,Chakraborti2010a}). The general aim was to study a many-agent statistical model
of closed economy (analogous to the kinetic theory model of ideal gases) \cite{Angle,Dragulescu2000a,Chakraborti2000a},
where $N$ agents exchange a quantity $x$, that
may be defined as wealth.
The states of agents are characterized by the wealth
$\{x_i\},~i=1,2,\dots,N$, and the total wealth $W=\sum_{i} x_i$ is conserved.
The question of interest is: ``What is the equilibrium distribution of wealth $f(x)$, such that $f(x) dx $ is the probability that in the steady state of the system, a randomly chosen agent will be found to have wealth between $x$ and $x + dx$?''

The evolution of the system is carried out according to a prescription,
which defines the trading rule between agents, where the agents 
interact with each other through a pair-wise interaction characterized
by a saving parameter $\lambda$, with $0 \le \lambda \le 1$. 
The dynamics of the model (CC)
is as follows
\cite{Chakraborti2000a}:
\begin{align}
  x_i' &= \lambda x_i + \epsilon (1-\lambda) (x_i + x_j) \, ,
  \nonumber \\
  x_j' &= \lambda x_j + (1-\epsilon) (1-\lambda) (x_i + x_j) \, .
  \label{sp1}
\end{align}
%corresponding to a $\Delta x$ in \eq{basic0} given by
%\begin{equation}
%  \Delta x =     (1 - \lambda) [ (1-\epsilon) x_i - \epsilon x_j ]  \, .
%\end{equation}
It can be noticed that in this way, the quantity $x$ is conserved
during the single transactions: $x_i'+x_j' = x_i + x_j$,
%(see \fg{fig:exchange}),
where $x_i'$ and $x_j'$ are the agent wealths
after the transaction has taken place.

This model for $ \lambda > 0 $ leads to an equilibrium distribution, with a mode $x_m>0$ and a zero limit for small $x$. For $ \lambda = 0 $, the model reproduces the results of Yakovenko \cite{Dragulescu2000a}, where the equilibrium distribution is the Gibb's distribution.
In general, the functional form for such distributions was conjectured to be a $\Gamma$-distribution on the basis of an analogy with the kinetic theory of gases: 
\begin{equation}
\label{gamman}
  f(x) = \frac{1}{\Gamma(n)} \left ( \frac{n}{\la x \ra}\right )^n  x^{n-1}
  \exp \left ( - \frac{nx}{\la x \ra} \right ) ,
\end{equation}
where
\begin{eqnarray}
  \label{D1}
  n
  = \frac{D(\lambda)}{2}
  = 1 + \frac{3\lambda}{1-\lambda}.%\frac{1 + 2\lambda}{1 - \lambda} \, .
  \label{n}
\end{eqnarray}
Indeed, starting from the Maxwell-Boltzmann distribution for the particle velocity in a $D$ dimensional gas, it can be shown that the equilibrium kinetic energy distribution coincides with the Gamma-distribution \eqref{gamman} with $n=\frac{D}{2}$. This conjecture is remarkably consistent with the fitting provided to numerical data \cite{Anirban2008,Patriarca2004a}.

As a further generalization, the agents could be assigned different saving propensities $\lambda_i$
\cite{Chatterjee2004a}.
In particular, uniformly distributed $\lambda_i$ in the interval $[0,1)$ had been studied numerically in Refs.~\cite{Chatterjee2004a}.
This model (CCM) is described by the trading rule
\begin{eqnarray}
  x_i' &=&
  \lambda_i x_i + \epsilon [ (1-\lambda_i) x_i + (1-\lambda_j) x_j ] \, ,
  \nonumber \\
  x_j' &=&
  \lambda_j x_j + (1-\epsilon) [(1-\lambda_i) x_i + (1-\lambda_j) x_j ] \, .
  \label{sp2}
\end{eqnarray}
One of the main features of this model, which is supported by theoretical considerations \cite{Mohanty,Repetowicz2005,Chakraborti2009}, is that the wealth distribution exhibits a robust power-law at large values of $x$,
\begin{equation}\label{f-power}
  f(x) \propto x^{-\alpha - 1} \, ,
\end{equation}
with a Pareto exponent $\alpha =1$ largely independent of the details of the $\lambda$-distribution. Note that other values of exponents can also be generated by modifying the exchange rules \cite{Arnab2007}.

%%%%%%%%%%%%%%%%%%%%%%%%%%%%%%%%%%%%%%%%%%%%%%%%%%%%%%%%%%%%%%%%%%%%%%%%%
\section{A Kinetic exchange model for opinion formation}
%%%%%%%%%%%%%%%%%%%%%%%%%%%%%%%%%%%%%%%%%%%%%%%%%%%%%%%%%%%%%%%%%%%%%%%%%
Toscani \cite{Toscani2006} had recently introduced and discussed kinetic models of (continuous) opinion formation involving both exchange of opinion between individual agents and diffusion of information. He showed that there are conditions which ensure that the kinetic model reaches non-trivial stationary states in case of lack of diffusion in correspondence of some opinion point, and obtained analytical results by considering a suitable asymptotic limit of the model yielding a Fokker-Planck equation for the distribution of opinion among individuals. Based on this model, During et al \cite{During2009} proposed another mathematical model for opinion formation in a society that is built of two groups, one group of ‘ordinary’ people and one group of ‘strong opinion leaders’. Starting from microscopic interactions among individuals, they arrived at a macroscopic description of the opinion formation process that is characterized by a system of Fokker–Planck-type equations. They discussed the steady states of the system, and extended it to incorporate emergence and decline of opinion leaders. On a different approach, 
Iniguez et al \cite{Iniguez2009} examined a
situation in which these non-identical individuals form their
opinions in information-transferring interactions with others. They
developed a dynamic network model, where they consider short range interactions for direct discussions
between pairs of individuals, long range interactions
for sensing the overall opinion modulated by the attitude
of an individual, and external field for outside influence.

%Finally, we believe that this very simple but elegant model inspired from the physics of kinetic theory can be easily adapted to model real-life scenarios and display rich social behaviour.

Following the CC and CCM models, described in the earlier section, we now propose a minimal model for the
collective dynamics of opinion $ O_i(t) $ of the $ i $-th person in the society of
$ N $ ($ N \longrightarrow \infty$) persons:

\begin{align}
  O_i(t+1) &= \lambda_i O_i(t) + \epsilon \lambda_j O_j(t) \, ,
  \nonumber \\
  O_j(t+1) &= \lambda_j O_j(t) + \epsilon' \lambda_i O_i(t) \, ,
  \label{op1}
\end{align}
where $-1 \le O_i(t) \le 1$ for all $i$ and $t$, and $0 \le \lambda_i\le 1$'s are \textit{quenched} variables (do not change with time, but vary from person to person), and $\epsilon$ and $\epsilon'$ are \textit{annealed} variables (change with time), that are random numbers uniformly distributed between 0 and 1.
\begin{figure}
\begin{center}
	\includegraphics[width=\linewidth]
        {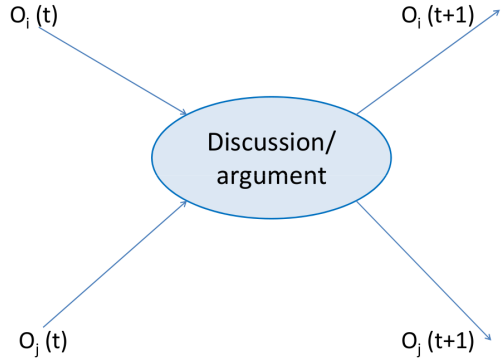}
    \end{center}
\caption{Schematic diagram of the minimal model where random discussions/arguments between two persons $ i $ and $ j $ with opinions $ O_i (t)$ and $ O_j (t)$, respectively, cause the update of opinions $ O_i (t+1)$ and $ O_j (t+1)$.}
\label{fig:exchange}
\end{figure}

The above described model dynamics, follows the two-body ``discussions/arguments'' modelled as \textit{scattering processes} and depicted schematically in \fg{fig:exchange}. It is based on the logic that during the discussion/argument event with any person $ j $, the person $ i $ with high/low ``conviction'' (parametrized by the $\lambda_i$), will \textit{retain} his/her own earlier opinion $O_i(t)$ proportional to the factor $\lambda_i$, and be \textit{influenced} to change the opinion by the $j$-th person's influence determined by a contribution which will depend on the $j$-th person's conviction $\lambda_j$ (and not by the factor $1 - \lambda_j$ as in market dynamics Eq. \ref{sp1} or \ref{sp2}).
Also, as \textit{no conservation in opinion is possible} (unlike in the market models above), the annealed variables $\epsilon$ and $\epsilon'$ are now considered to be \textit{uncorrelated}. Additionally we assume that $| O_i(t) | \leq 1$, for all $ i $ and $ t $.

\subsection{Homogeneous conviction factor case}

When we assume $\lambda_i = \lambda$ for all $i$ (equivalent to the CC model for market dynamics), the above equations reduce to
\begin{align}
  O_i(t+1) &= \lambda (O_i(t) + \epsilon O_j(t)) \, ,
  \nonumber \\
  O_j(t+1) &= \lambda (O_j(t) + \epsilon' O_i(t)) \, .
  \label{op2}
\end{align}
This leads to an intriguing \textit{spontaneous symmetry breaking} transition beyond a threshold value of $\lambda_c = 2/3$. Specifically, following the above dynamics, starting from random (drawn uniformly) positive and negative values of $O_i(0)$ (at $t = 0$) (``symmetric'' state, when the \textit{order parameter} $\la O\ra \equiv (1/N)\sum_i O_i(t=0) = 0$), leads the system to
collectively evolving to two kinds of state: 
\begin{enumerate}
\item ``Para'' or ``indifferent'' state, where $O_i(t)$'s are \textit{all} zeros ($\la O\ra = 0$) after a ``relaxation'' time $\tau$, for $\lambda$ values less than $\lambda_c=2/3$; or 
\item ``Symmetry broken''  or ``polarised'' state, where $O_i(t)$'s are either all positive or  all negative ($\la O\ra \ne 0$) after a ``relaxation'' time $\tau$, for $\lambda > 2/3$.
\end{enumerate} 
One can easily see that for $\lambda$ values less than $2/3$, with $\la\epsilon\ra = 1/2$, the recursion relation for the \textit{order parameter} $\la O\ra$ becomes a simple multipler equation with the value of the multiplier less than unity, leading to $\la O\ra = 0$ eventually. For higher values of $\lambda$, the fluctuations in $\epsilon$ are important
(and cannot be replaced by its simple average, as above) because of asymmetric contributions from the second term of
both the above equations 
%(if $|O(t+1)|$ becomes $> 1$, partial contribution of the second term is accepted, while for lower contributions are accepted entirely). 
(if the contribution of the second term in Eq. \ref{op2} takes the value of $|O(t+1)|$ to greater than unity, only partial contribution of the second term is accepted, while for its lower values the acceptance is full). 
We find, this seemingly leads to a \textit{discontinous} or ``first order'' symmetry breaking transition at $\lambda_c =2/3$ (see \fg{fig:phase}). The details of this transition will
be reported elsewhere \cite{Mehdi}.

\begin{figure}
\begin{center}
	\includegraphics[width=0.8\linewidth]
        {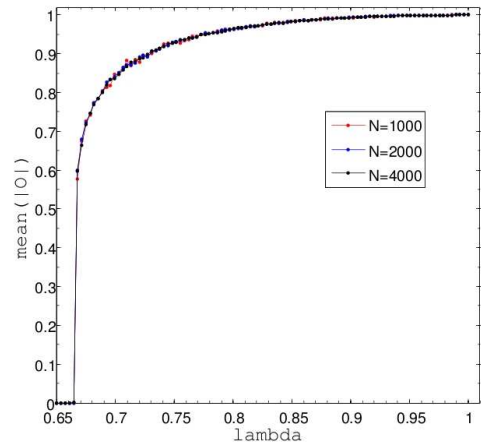}
    \end{center}
\caption{The variation of the order parameter $\la O\ra \equiv (1/N)\sum_i O_i(t)$ against $\lambda$.}
\label{fig:phase}
\end{figure}

\subsection{Heterogeneous conviction factor case}

Here, we assume $\lambda_i$'s to be uniformly spread in the interval [0,1) (equivalent to the CCM model for market
dynamics). We study similarly, starting from ``symmetric'' states (with random positive and negative values of $O_i(0)$, the evolution of the system. The dynamics here leads collectively to the ``Polarized'' or ``Symmetry broken" state ($O_i(t)$ are either all positive or all negative, for all $i$, and times $t > \tau$ ) only.
The ``indifferent'' states (with $O_i(t) = 0$ for all $i$, for times $t > \tau$) disappear in the large system size limit, although this is clearly a \textit{fixed point} of the dynamics given by Eq. \ref{op1}. We believe, this is also a clear feature of the opinion dynamics model proposed by Iniguez et al \cite{Iniguez2009}, where also this state is surely a fixed point of their model.

It may be noted that the above dynamics can be considerably modified by the presence of ``polarizing field'' terms $ h_i $ (fixed over time $ t $ but dependent on person $ i $), added linearly to the dynamical equations Eq. \ref{op1} of $ O_i(t)$. Such ``fields'' can be provided by the ``influences'' of the media in the society. Detailed analyses of the field terms, etc. will be reported elsewhere \cite{Lallouache2010b}.

\section{Discussion and Summary}

The appearance of spontaneous symmetry breaking in this kinetic opinion exchange model is truely remarkable. It
appears to be one of the simplest collective dynamical model of many-body dynamics showing non-trivial phase
transition behaviour. The details of this transition is under investigation and will be reported elsewhere.

\begin{acknowledgments}
%\textbf{Acknowledgments}
The authors acknowledge F. Abergel, A.S. Chakrabarti, A. Chatterjee, A. Jedidi, K. Kaski and N. Millot for useful discussions and comments.
\end{acknowledgments}

%%%%%%%%%%%%%%%%%%%%%%%%%%%%%%%%%%%%%%%%%%%%%%%%%%%%%%%%%%%%%%%%%%%%%%%%%

\end{document}